\documentstyle[aps,preprint]{revtex}
\begin{document}
\draft

\title{\large Highly deformed $^{40}$Ca configurations in $^{28}$Si + $^{12}$C}

\author{M.~Rousseau\thanks{Corresponding author. Electronic address:
marc.rousseau@ires.in2p3.fr}, C.~Beck, C.~Bhattacharya\thanks{Permanent
address: VECC, 1/AF Bidhan Nagar, Kolkata 64, India.}, V.~Rauch, O.~Dorvaux,
K.~Eddahbi, C.~Enaux, R.M.~Freeman, F.~Haas, D.~Mahboub{\thanks{Present
address: University of Surrey, Guildford GU2 7XH, United Kingdom.}}, R. 
Nouicer\thanks{Present address: Department of Physics, University of Illinois
at Chicago, Chicago, Illinois 60607-7059, USA.}, P.~Papka,
O.~Stezowski{\thanks{Permanent address: IPN Lyon, F-69622 Villeurbanne,
France.}}, and S.~Szilner } 

\address{\it Institut de Recherches Subatomiques, UMR7500, Institut National de
Physique Nucl\'eaire et de Physique des Particules - Centre National de la
Recherche Scientifique/Universit\'e Louis Pasteur, 23 rue du Loess, B.P. 28,
F-67037 Strasbourg Cedex 2, France}

\author{A.~Hachem and E.~Martin}

\address{\it Universit\'e de Nice-Sophia Antipolis, F-06108 Nice, France}

\author{S.J.~Sanders and A.K.~Dummer{\thanks{Present address: Triangle
Universities Nuclear Laboratory, University of North Carolina, Durham, NC
27708-0308, USA}}} 

\address{\it Department of Physics and Astronomy, University of Kansas,
Lawrence, Kansas 66045, USA}

\author{A.~Szanto de Toledo}

\address{\it Departamento de F\'{\i}sica Nuclear, Instituto de F\'{\i}sica da
Universidade de S\~ao Paulo, C.P. 66318-05315-970 - S\~ao Paulo, Brazil} 

\date{\today}
\maketitle

\newpage

\begin{abstract}

{The possible occurrence of highly deformed configurations in the $^{40}$Ca
di-nuclear system formed in the $^{28}$Si + $^{12}$C reaction is investigated
by analyzing the spectra of emitted light charged particles. Both inclusive and
exclusive measurements of the heavy fragments (A $\geq$ 10) and their
associated light charged particles (protons and $\alpha$ particles) have been
made at the IReS Strasbourg {\sc VIVITRON} Tandem facility at bombarding
energies of $E_{lab}$($^{28}$Si) = 112 MeV and 180 MeV by using the {\sc ICARE}
charged particle multidetector array. The energy spectra, velocity
distributions, and both in-plane and out-of-plane angular correlations of light
charged particles are compared to statistical-model calculations using a
consistent set of parameters with spin-dependent level densities. The analysis
suggests the onset of large nuclear deformation in $^{40}$Ca at high spin.}

\end{abstract}
{PACS number(s): 25.70.Gh, 25.70.Jj, 25.70.Mn, 24.60.Dr}

\newpage

\section{INTRODUCTION}

The formation and binary decay of light nuclear systems in the A$_{CN}$
$\leq$ 60 mass region produced by low-energy (E$_{lab}$ $\leq$ 7 MeV/nucleon)
heavy-ion reactions has been extensively studied both from the experimental and
the theoretical points of view \cite{Sanders99}. In most of the reactions
studied the binary breakup of the compound nucleus (CN) is seen as either a
fusion-fission (FF) \cite{Sanders99,Matsuse97} or a deep-inelastic (DI)
orbiting \cite{Shivakumar87} process. The large-angle orbiting yields are
found to be particularly strong in the $^{28}$Si + $^{12}$C reaction
\cite{Shapira82}, as illustrated by Fig.~1 which summarizes some of the
experimental results that have been collected for this system; i.e., orbiting
cross sections~\cite{Shapira82,Shapira84} and total evaporation residue (ER)
cross
sections~\cite{Gary82,Lesko82,Nagashima82,Harmon86,Harmon88,Vineyard93,Arena94}.
Since many of the conjectured features for orbiting yields are similar to those
expected for the FF mechanism, it is difficult to fully discount FF as a
possible explanation for the large energy-damped $^{28}$Si + $^{12}$C
yields~\cite{Shivakumar87,Shapira82,Shapira84}. However, FF
calculations~\cite{Sanders99} significantly underpredict the cross sections
measured in the carbon channel by almost a factor of 3, thus suggesting an
alternative mechanism (see Fig.~1). FF, DI orbiting, and even
molecular-resonance behavior may all be active~\cite{Pocanic85} in the
large-angle yields of the $^{28}$Si + $^{12}$C
reaction~\cite{Ost79,Barrette79}. The back-angle elastic scattering of
$^{28}$Si ions from $^{12}$C displays structured excitation functions and
oscillatory angular distributions in agreement with the relatively weak
absorption of this system \cite{Beck94}. Moreover, the resonant gross
structure~\cite{Ost79} is fragmented into very striking intermediate width
resonant structure~\cite{Barrette79}. 

Superdeformed (SD) rotational bands have been found in various mass regions (A
= 60, 80, 130, 150 and 190) and, very recently, SD bands have also been
discovered in the N = Z nuclei $^{36}$Ar~\cite{Svensson00,Svensson01} and
$^{40}$Ca~\cite{Ideguchi01}. These new results make the A$_{CN}$ $\approx$ 40
mass region of particular interest since quasimolecular resonances have also
been observed in both the $^{36}$Ar and $^{40}$Ca dinuclear
systems~\cite{Pocanic85}. Although there is no experimental evidence to link
the SD bands with the higher lying rotational bands formed by known
quasimolecular resonances, both phenomena are believed to originate from highly
deformed configurations of these systems. Since the detection of light charged
particles (LCP) is relatively simple, the analysis of their spectral shapes is
another good tool for exploring nuclear deformation and other properties of hot
rotating nuclei at high angular momenta. Experimental evidence for
angular-momentum-dependent spectral shapes has already been extensively
discussed in the 
literature~\cite{Choudhury84,Majka87,Govil87,Fornal88,Viesti88,Fornal89,Larana89,Huizenga89,Fornal91a,Agnihotri93,Govil98,Bandyopadhyay99,Govil00a,Govil00b,Bandy01}
and, in particular, the $^{24}$Mg + $^{16}$O reaction~\cite{Fornal91b}, which
reaches the $^{40}$Ca CN, has been studied in detail. Strong deformation
effects have also been deduced from angular correlation data for the fusion
reaction $^{28}$Si($^{12}$C,2$\alpha$)$^{32}$S$_{g.s.}$ at E$_{lab}$ = 70 
MeV~\cite{Alamanos83}. We decided to investigate the $^{40}$Ca nucleus produced
through the $^{28}$Si + $^{12}$C reaction at beam energies of E$_{lab}$ = 112
MeV and 180 MeV. As can be observed in Fig.~1, model calculations suggest that 
at the lowest incident energy
of the present work the orbiting process is dominant for the C and O channels
whereas at E$_{lab}$ = 180 MeV a large part of the O and N fully-damped yields
may also result from a FF mechanism. In this article we will focus on the LCP's
found in coincidence with heavy fragments.

The present paper is organized in the following way. Sec.~II describes the
experimental procedures and the data analysis. Sec.~III presents the inclusive
and the exclusive $^{28}$Si + $^{12}$C data (part of the experimental results
presented here in detail have already been briefly reported elsewhere
~\cite{Rousseau00,Beck00,Rousseau01a,Rousseau01b,Bhattacharya01}). The data are
analyzed using the Hauser-Feshbach evaporation code {\sc
CACARIZO}~\cite{Choudhury84,Majka87,Viesti88} using a consistent set of
parameters which has been found to successfully reproduce $^{24}$Mg + $^{16}$O
reaction results~\cite{Fornal91b}. The full statistical-model calculations,
using Monte Carlo techniques to account for the experimental acceptance when
comparing to the experimental exclusive data, are presented in Sec.~IV. The
strong cluster emission of $^{8}$Be which is observed by the experiment is also
discussed in this section. We end with a summary of our conclusion in Sec.~V. 

\bigskip

\section{EXPERIMENTAL PROCEDURES AND DATA ANALYSIS}

The experiments were performed at the {\sc VIVITRON} Tandem facility of the
IReS Strasbourg laboratory using 112 MeV and 180 MeV $^{28}$Si beams which were
incident on $^{12}$C targets (160 and 180 $\mu$g/cm${^2}$ thick, respectively)
mounted in the ICARE scattering chamber~\cite{Belier94,Bellot97}. The effective
thicknesses of the $^{12}$C targets were accurately determined using Rutherford
back scattering (RBS) techniques with $^{1}$H and $^{4}$He beams provided by
the Strasbourg 4 MV Van de Graaff
accelerator~\cite{Rousseau01a,Bhattacharya99,Bhattacharya02}. The carbon
buildup corrections were found to be less than 2$ \% $ of the total of C atoms
in the targets. Both the heavy fragments (A $\geq$ 10) and their associated
LCP's (protons and $\alpha$ particles) were detected in coincidence using the
{\sc ICARE} charged-particle multidetector array \cite{Belier94,Bellot97} which
consists of nearly 40 telescopes. Inclusive data have also been collected for 
heavy fragments and LCP's. 

The setup of the measurement at E$_{lab}$($^{28}$Si) = 112 MeV was designed to
collect only in-plane coincidences, whereas the setup at E$_{lab}$($^{28}$Si) =
180 MeV allowed both in-plane and out-of-plane angular correlations to be
measured (see Table I). The heavy fragments consisting of ER as well as
quasi-elastic, deep-inelastic, and fusion-fission fragments were detected in 8
gas-silicon hybrid telescopes (IC), each composed of a 4.8 cm thick ionization
chamber, with a thin Mylar entrance window, followed by a 500 $\mu$m thick
Si(SB) detector. The IC's were located at forward angles in two
distinct reaction planes (for each plane, the positive and negative angles in
Table I are defined in a consistent manner as for the LCP detectors described
below). The in-plane detection of coincident LCP's was done using 4
three-element telescopes (TL3) (40 $\mu$m Si, 300 $\mu$m Si, and 2 cm CsI(Tl))
placed at forward angles, 16 two-element telescopes (TL2) (40 $\mu$m Si, 2 cm
CsI(Tl)) placed at forward and backward angles and, finally, two other IC
telescopes located at the most backward angles (see Table I). The CsI(Tl)
scintillators were coupled to photodiode readouts. The IC's were filled with
isobutane at a pressure of 30 Torr for the backward angle telescopes and of 60
Torr for the forward angle detectors, thus allowing for the simultaneous
measurement of both light and heavy fragments. 

For the measurement at E$_{lab}$($^{28}$Si) = 180 MeV, two distinct reaction
planes were defined as shown in Table I. One for in-plane correlations and a
second one, perpendicular to the LCP detection plane, for out-of-plane
correlation measurements. The heavy fragments were detected in 10 IC's located
at forward angles for both the in-plane coincidences and out-of-plane
coincidences. Both the in-plane and out-of-plane coincident LCP's were detected
using 3 TL3's placed at forward angles and 24 TL2's placed at forward and
backward angles. The IC's were filled with isobutane at a pressure of 60 Torr. 

The acceptance of each telescope was defined by thick aluminum collimators.
The distances of these telescopes from the target ranged from 10.0 to 30.0 cm,
and the solid angles varied from 1.0 msr at the most forward angles to 5.0 msr
at the backward angles, according to the expected counting rates. 

The energy calibrations of the different telescopes of the {\sc ICARE}
multidetector array were done using radioactive $^{228}$Th and $^{241}$Am
$\alpha$-particle sources in the 5-9 MeV energy range, a precision pulser, and
elastic scatterings of 112 MeV and 180 MeV $^{28}$Si from $^{197}$Au,
$^{28}$Si, and $^{12}$C targets in a standard manner. In addition, the
$^{12}$C($^{16}$O,$\alpha$)$^{24}$Mg$^{*}$ reaction at E$_{lab}$ = 53 MeV
\cite{Rousseau01a} was used to provide known energies of $\alpha$ particles 
feeding the $^{24}$Mg excited states, thus allowing for calibration of the
backward angle detectors. The proton calibration was achieved using scattered
protons from C$_{5}$H$_{4}$O$_{2}$ Formwar (polyvinyl formal) targets bombarded
in reverse kinematics reactions with both $^{28}$Si and $^{16}$O beams. On an
event-by-event basis, corrections were applied for energy loss of heavy
fragments (A $\geq$ 10) in the targets and in the entrance window Mylar foils
of the IC's and thin Al-Mylar foils of the Si diodes. A correction was also
applied for the pulse height defect in the Si detectors. The IC energy
thresholds and energy resolution for heavy fragments are better than 1.5
MeV/nucleon and 0.7$\%$, respectively. The total energy resolution of 8.78 MeV
$\alpha$ particles from thorium sources has been found to be better than
2.2$\%$ for both the three-element and two-element light-ion CsI(Tl)
telescopes. Absolute cross sections of inclusive measurements could be obtained
within 10-12$\%$ uncertainties, with 3-5$\%$ uncertainties in the target
thickness and to 8-10$\%$ uncertainties in the electronic deadtime corrections.
More details on the experimental setup of {\sc ICARE} and on the analysis
procedures can be found in
Refs.~\cite{Bhattacharya99,Bhattacharya02,Beck00,Rousseau01a,Rousseau01b} and
references therein. 

\bigskip

\section{EXPERIMENTAL RESULTS}

\subsection{Inclusive data}

The fragments with Z = 15-17 have inclusive energy spectra that are typical of
ER's. The inclusive data of the binary fragments with Z = 5-14 were obtained
by integrating their 1/sin$\theta_{c.m.}$ angular distributions. In particular
the results of the cross sections with Z = 6-8 are compared with the previously
measured excitation functions~\cite{Shapira82,Shapira84} in Fig.~1. The
agreement between both sets of data is satisfactory within the error bars. 

The LCP inclusive data can shed some light on the reaction mechanism. Typical
inclusive energy spectra of $\alpha$ particles are shown for the $^{28}$Si +
$^{12}$C reaction at E$_{lab}$ = 112 MeV in Figs.~2-(a) and 2-(b) and at E$_{lab}$
= 180 MeV in Figs.~2-(c) and 2-(d) at the indicated angles. The solid points
(with error bars visible when greater than the size of the points) are the 
experimental data whereas the solid and dashed lines are statistical model
calculations discussed in Sec.~IV.B. The $\alpha$-particle energy spectra have
a Maxwellian shape with an exponential fall-off at high energy which reflects a
relatively high effective temperature (T$_{slope}$ $\approx$ [8E$^{*}_{\small
CN}$/A$_{\small CN}$]$^{1/2}$ = 3.67 MeV for E$_{lab}$ = 180 MeV) of the
decaying nucleus. The spectral shape and high-energy slopes are also found to
be essentially independent of angle in the c.m. system. These observations
suggest a statistical de-excitation process arising from a thermalized source
such as the $^{40}$Ca CN and are consistent with a previous study at E$_{lab}$
= 150 MeV~\cite{Kildir95}. 
  
The velocity contour maps of the LCP Galilean-invariant differential cross
sections (d$^{2}\sigma$/d$\Omega$dE)p$^{-1}$c$^{-1}$ as a function of the LCP
velocity are known to provide an overall picture of the reaction pattern. From
this pattern the velocity of the emission source can be determined in order to
better characterize the nature of the reaction mechanism. Such typical plots
~\cite{Rousseau00,Rousseau01a,Rousseau01b,Kildir95} (not shown here) can
again be understood by assuming a sequential evaporative process and successive
emission sources starting with the thermally equilibrated $^{40}$Ca$^{*}$ CN
and ending with the final source characterized by a complete freeze-out of the
residual nucleus. The Galilean-invariant cross section contours form arcs that
are centered at V$_{c.m.}$~\cite{Kildir95}, as expected for a
complete fusion-evaporation (CF) mechanism followed by isotropic evaporation. 

\subsection{Exclusive data}

\subsubsection{LCP energy spectra}

The exclusive LCP events are also largely consistent with a CF mechanism with
the light particles being in coincidence with ER's with Z = 15-17. The data
taken with the IC's located at more backward angles (larger than -15$^{\circ}$)
are not considered in the following analysis since the statistics for
fusion-evaporation events are too low. The experimental data of Figs. 3 and 4
are given by the solid points, with error bars visible when greater than the
size of the points. At E$_{lab}$ = 180 MeV, the spectral shapes of $\alpha$ particles in coincidence
with P (Z=15) ER's, shown in Fig.~4, are very similar to the inclusive energy
spectra of Fig.~2. On the other hand the energy spectra of $\alpha$ particles
in coincidence with S (Z=16) ER's are more complicated as they show other
non-evaporative sub-structures (their angular dependence are indicative of a
binary decay origin) which are superimposed on the ``statistical" Maxwellian
shape. These additional non-statistical components will be discussed as arising
from a $^{8}$Be cluster emission in Sec.~IV.C. At E$_{lab}$ = 112 MeV, shown in 
Fig.~3, for Z = 15 the
high-energy components showing up at the large angles may arise from $\alpha$,
3p evaporation cascades. 

Fig.~5 presents the corresponding $^{28}$Si + $^{12}$C exclusive energy spectra
of protons emitted in coincidence with individual ER's (Z = 15 and 16) at
E$_{lab}$ = 180 MeV. Their spectral shapes are also Maxwellian with the typical
exponential fall-off at high energy, characteristic of a statistical CN decay
process. 

\subsubsection{In-plane angular correlations}

The in-plane angular correlations of $\alpha$ particles and protons in
coincidence with ER's produced in the $^{28}$Si + $^{12}$C reaction, are shown
in Figs.~6 and 7 at E$_{lab}$ = 112 MeV and 180 MeV, respectively. The angular
correlations are peaked strongly on the opposite side of the beam direction
from the ER detectors which were located at $\Theta^{ER}_{lab}$ = -15$^{\circ}$
or $\Theta^{ER}_{lab}$ = -10$^{\circ}$ for the two energies, respectively. The
observed peaking of the LCP yields on the opposite side of the beam from the IC
is the result of momentum conservation. The angular correlations of both the
protons and $\alpha$ particles show the same behavior for all ER's.

\subsubsection{Out-of-plane angular correlations}

Fig.~8 displays the out-of-plane angular correlations of $\alpha$ particles
(circles) and protons (triangles) in coincidence with individual ER's detected
at $\Theta_{lab}$ = 10$^\circ$, produced in the $^{28}$Si(180 MeV) + $^{12}$C
reaction. The angular distributions have a behavior following a typical
exp(-a$sin^2$($\theta_{lab}$))
shape~\cite{Rousseau01a,Beck95,Mahboub96,Mahboub02}, with possibly two
components visible for $\alpha$ particles in coincidence with Z = 14, 15 and
16, plus a broadening of the angular correlations at backward angles. This
broadening effect may result from $\alpha$ particles being able to be emitted
at the beginning or at the end of the decay chain, where the angular momentum
becomes smaller towards the end of the chain. As the protons cannot remove as
much angular momentum as do the $\alpha$ particles the broadening effect is
less significant in the proton angular correlation. The solid lines shown in
the figure are the results of statistical-model predictions for CF and
equilibrium decay using the Monte Carlo evaporation code {\sc
CACARIZO}~\cite{Choudhury84,Majka87,Viesti88}, as discussed in the next
Section. 

\bigskip

\section{DISCUSSION}

\subsection{Statistical-model calculations}

The evaporation of light particles from a highly excited CN is a well known
decay process up to very high excitation energies and
spins~\cite{Stokstad85,Cole00,Charity00}. The interpretation of LCP data
requires a careful treatment of the light particle emission properties in the
statistical-model description. Most of the available statistical-model computer
codes~\cite{Sanders99,Matsuse97,Stokstad85,Cole00,Charity00} are based on the
Hauser-Feshbach formalism and are able to follow the CN decay by a cascade of
evaporated LCP's and neutrons. In particular, a detailed analysis of the
exclusive data can be undertaken by the use of Monte Carlo
versions~\cite{Cole00} of some of these statistical-model codes in which the
filtering of the events can reproduce the experimental conditions. The
statistical-model analysis of the present data has been performed using the
Hauser-Feshbach evaporation code {\sc CACARIZO}~\cite{Choudhury84}. {\sc
CACARIZO} is a Monte Carlo version of the statistical-model code {\sc
CASCADE}~\cite{Puhlhofer77}, which has evolved with many modifications and
extensions~\cite{Choudhury84,Majka87,Viesti88} from the original code. In this
program the effective experimental geometry of the {\sc ICARE} detectors is
properly taken into account. It is assumed that a single CN is created with a
well defined excitation energy and angular momentum distribution, and the
de-excitation chain is followed step by step and recorded as an event file. The
generated events are then analyzed using a subsequent filtering code {\sc
ANALYSIS}~\cite{Majka87} in which the locations and the solid angles of all the
{\sc ICARE} telescopes are explicitly specified. This program allows the
determination of the different types of events of interest. Such events can be
sorted (singles events, coincidence events, etc.) and the corresponding
particle spectra and angular distributions can be created. 

The CN  angular momentum distributions needed as the primary input for the
calculations were specified using the critical angular momentum for fusion
L$_{crit}$ and the diffuseness parameter $\Delta$L. They were taken from ER
cross section data compiled for the $^{28}$Si + $^{12}$C fusion process by
Vineyard {\it et al.}~\cite{Vineyard93}, without including fission competition.
A fixed value of $\Delta$L = 1$\hbar$ (optimized at low energy by a previous
statistical-model analysis of this reaction~\cite{Alamanos83}) was assumed for
the calculations. It has been checked that the calculated spectra are not
sensitive to slight changes in the critical angular momentum or to explicit
inclusion of the fission competition. The parameter sets used for the
calculations are summarized in Table II. 

The other standard ingredients for statistical-model calculations are the
formulations of the nuclear level densities and of the barrier transmission
probabilities. The transmission coefficients were derived from Optical Model
(OM) calculations using potential parameters of light particle induced
reactions deduced by Wilmore and Hodgson~\cite{Wilmore64}, Perey and
Perey~\cite{Perey63}, and Huizenga~\cite{Huizenga61} for the neutrons, protons
and $\alpha$ particles, respectively. For spin regions where the standard
rotating liquid drop model (RLDM)~\cite{Cohen74} as well as the finite-range
liquid drop model (FRLDM)~\cite{Sierk86} still predict essentially spherical
shapes, these sets of transmission coefficients have been found adequate in the
considered mass region. However, in recent years it has been observed that when
the angular momentum is increased to values for which FRLDM predicts
significant deformations, statistical-model calculations using such standard
parameters cannot always predict satisfactorily the shape of the evaporated
$\alpha$-particle energy spectra 
\cite{Choudhury84,Majka87,Govil87,Fornal88,Viesti88,Fornal89,Larana89,Huizenga89,Fornal91a,Agnihotri93,Govil98,Bandyopadhyay99,Govil00a,Govil00b,Bandy01}.
The calculated average energies of the $\alpha$ particles are found to be much
higher than the corresponding experimental results. Several attempts have been
made to explain this anomaly either by changing the emission barrier or by
using spin-dependent level densities. Adjusting the emission barriers and
corresponding transmission probabilities affects the lower-energy part of the
calculated evaporation spectra. On the other hand the high-energy part of the
spectra depends crucially on the available phase space obtained from the level
densities at high spin. In hot rotating nuclei formed in heavy-ion reactions,
the energy level density at higher angular momentum is spin dependent. The
level density, $\rho(E,J)$, for a given angular momentum $J$ and energy $E$ is
given by the well known Fermi gas expression with equidistant single-particle
levels and a constant level density parameter $a$: 

\begin{equation}
\rho(E,J) = {\frac{(2J+1)}{12}}a^{1/2}
           ({\frac{ \hbar^2}{2 {\cal J}_{eff}}}) ^{3/2}
           {\frac{1}{(E-\Delta-T-E_J)^2} }$\rm exp$(2[a(E-\Delta-T-E_J)]^{1/2})
\label{lev}
\end{equation}
\noindent
where T is the ``nuclear" temperature and $\Delta$ is the pairing
correction~\cite{Dilg73}. The quantity E$_J$ = $\frac{ \hbar^2}{2 {\cal
J}_{eff}}$J(J+1) is the rotational energy, with ${\cal J}_{eff} = {\cal J}_0
\times (1+\delta_1J^2+\delta_2J^4)$ being the effective moment of inertia,
where ${\cal J}_0$ at high excitation energy and high angular momentum is
considered to be the rigid body moment of inertia and $\delta_1$ and $\delta_2$
are the ``deformability parameters" 
\cite{Choudhury84,Majka87,Govil87,Viesti88,Huizenga89,Fornal91a,Agnihotri93,Govil98,Bandyopadhyay99,Govil00a,Govil00b,Bandy01}.

The level density parameter is constant and is set equal to $a$ = A/8
MeV$^{-1}$, a value which is in agreement with previous works 
\cite{Fornal91b,Shlomo91,Toke81}. In principle, the value of $a$ may be
affected by dynamical deformation: rotation induces rearrangement of the
single-particle level scheme and the altered nuclear surface area
~\cite{Huizenga89} affects the macroscopic energy of the system. The $a$
parameter becomes more important when the nuclear deformation
increases~\cite{Toke81}. However, in the present work we assume a constant
value and rather introduce deformation effects through the deformability
parameters. A constant value of $a$ = A/8 is in agreement with various
authors~\cite{Huizenga89,Fornal91b}, as well as with theoretical studies by
Shlomo and Natowitz~\cite{Shlomo91}, by T\"oke and Swiatecki~\cite{Toke81},
and with experimental results obtained very recently in the A$_{CN}$ = 60 mass
region~\cite{Janker99}. 

No attempt was made to modify the transmission coefficients since it has been
shown that the effective barrier heights are fairly insensitive to the nuclear
deformation~\cite{Huizenga89}. On the other hand, by changing the deformability
parameters $\delta_1$ and $\delta_2$ one can simulate the spin-dependent level
density~\cite{Govil87,Viesti88,Huizenga89,Agnihotri93} associated with a larger
nuclear deformation, and thus better reproduce the experimental data. 

\subsection{Deformation effects in $^{40}$Ca}

In the present analysis, following the procedure proposed by Huizenga {\it et
al.}~\cite{Huizenga89}, we empirically modify the phase space open to
statistical decay by lowering the Yrast line with adjustment of the
deformability parameters so as to fit the available experimental
data~\cite{Govil87,Viesti88}. We may also take into account the fact that the
deformation should be attenuated during the subsequent emission processes:
i.e., there is a readjustment of shape of the nascent final nucleus and a
change of collective to intrinsic excitation during the particle-evaporation
process. Such an analysis was suggested earlier by Blann and
Komoto~\cite{Blann81}, but with the assumption that the deformation is a frozen
degree of freedom through the decay chain. Dynamical effects related to the
shape relaxation during the de-excitation process have been incorporated into
statistical-model codes~\cite{Fornal89,Fornal91a}. For the CACARIZO
calculations done here, it is assumed that memory of formation details are lost
after each step, with only the conserved quantities such as total energy and
spin preserved during the decay sequence. The {\sc CACARIZO} calculations have
been performed using two sets of input parameters: the first one with standard
liquid drop parameters (parameter {\bf set A}), consistent with the deformation
of RLDM~\cite{Cohen74} and of FRLDM with finite-range corrections of
Sierk~\cite{Sierk86}, and the second one with larger values for the
deformability parameters
\cite{Bhattacharya99,Bhattacharya02,Rousseau00,Beck00,Bhattacharya01}
(parameter {\bf set B}) which are listed in Table \ref{table3}. In the
FRLDM~\cite{Sierk86} the CN can be considered as spherical or slightly deformed
at low bombarding energy, becoming strongly deformed as the spin increases. 
 
The dashed lines in Fig.~2 show the predictions of {\sc CACARIZO} for $^{28}$Si
+ $^{12}$C at both bombarding energies using the parameter {\bf set A}
consistent with FRLDM deformation~\cite{Sierk86}. It can be observed that the
average energies of the measured $\alpha$-particle inclusive spectra are lower
than those predicted by these statistical-model calculations. The solid lines
of Fig.~2 show the predictions of {\sc CACARIZO} using the same increased
values of the spin deformation parameters (see parameter {\bf set B} given in
Table II) for both energies, and the agreement is considerably improved. 

The exclusive energy spectra of the $\alpha$ particles in coincidence with
individual ER's (Z = 15 and Z = 16) are displayed in Figs.~3 and 4 for the two
bombarding energies E$_{lab}$ =  112 MeV and 180 MeV, respectively. It can be
observed that the spectra in coincidence with the P residues are well
reproduced by using the deformation
effects~\cite{Bhattacharya99,Rousseau00,Beck00,Bhattacharya01}. The solid lines
in Figs.~3 and 4 show the predictions of {\sc CACARIZO} using the parameter
{\bf set B} with $\delta_1$ = 2.5 x 10$^{-4}$ and $\delta_2$ = 5.0 x 10$^{-7}$
chosen to reproduce the data consistently at the two bombarding energies. On
the other hand, by using the standard liquid drop deformability parameter {\bf
set A} with no extra deformation (i.e. with small values of $\delta _1$ and
$\delta_2$), the observed average energies from the exclusive
$\alpha$-particles spectra are, as found for the inclusive data, lower than
those predicted \cite{Rousseau01a} by the statistical model. In this case the
{\sc CACARIZO} parameters are similar to the standard parameters used in a
previous study of the 130 MeV $^{16}$O +$^{24}$Mg reaction \cite{Fornal91b},
with the use of the angular momentum dependent level densities. 

On inspection of Figs.~3 and 4 a large difference can be noticed when
comparing the calculated and experimental energy spectra associated with S
residues and those associated with P ER's~\cite{Beck00}. The latter are
reasonably well reproduced by the {\sc CACARIZO} calculations, whereas the
model does not predict the shape of the spectra obtained in coincidence with S
residues at backward angles ($\theta_{\alpha}\geq 40^{\circ}$ at E$_{lab}$ =
112 MeV and $\theta_{\alpha}\geq 70^{\circ}$ at E$_{lab}$ = 180 MeV). An
additional, non-statistical, $\alpha$-particle  component is suggested in
Section IV.C to arise from a $^{8}$Be emission process. This is consistent with
the discrepancies also observed at backward angles in the in-plane angular
correlations of Fig.~6. For both $\alpha$ particles and protons the
calculations significantly underpredict the yields at the negative angles at
E$_{lab}$ = 112 MeV. However the agreement is more satisfactory for protons at
E$_{lab}$ = 180 MeV as shown in Fig.~7. 

As shown in Fig.~5 {\sc CACARIZO} calculations are also able to reproduce
the shape of exclusive proton spectra for the $^{28}$Si(180 MeV) + $^{12}$C
reaction. Compared to the $\alpha$ particles, it may be mentioned that the
energy spectra of the protons do not shift as significantly as the
spin-dependent parametrization of the moment of inertia is introduced. The
statistical-model results using the two different parameter sets reproduce
equally well the experimental velocity spectra and angular correlations. The
statistical-model calculations displayed for protons in Fig.~5 have been
performed with parameter {\bf set B} (solid lines) including the deformation
effects (calculated curves with parameter {\bf set A} are indistinguishable
from the solid lines). 

In order to better determine the magnitude of the influence of deformation
effects in the CN and the residual nuclei which are suggested by our choice of
statistical-model approach, we have proposed a very simple
procedure~\cite{Bhattacharya02,Beck95,Mahboub96,Mahboub02}. The effective
moment of inertia is 
expressed as 
${\cal J}_{eff}$ = ${\frac{2}{5}}$MR$^{2}$ = ${\frac{1}{5}}$M(b$^{2}$+a$^{2}$)
with the volume conservation condition: V = ${\frac{4}{3}}\pi$abc, 
where b and a are the major and minor axis, and c is the rotational axis
of the CN.
In the case of an oblate shape a = b and ${\cal J}_{eff}$ =
${\frac{2}{5}}$Ma$^{2}$ 
and V = ${\frac{4}{3}}\pi$a$^{2}$c. The axis ratio is equal to 
$\delta$ = a/c = (1+$\delta_1$J$^{2}$+$\delta_2$J$^{4}$)$^{3/2}$.
In the case of a prolate shape a = c and ${\cal J}_0$ =
${\frac{1}{5}}$M(b$^{2}$+a$^{2}$) and  V = ${\frac{4}{3}}\pi$a$^{2}$b. We
obtain the equation: 1+(3-$\gamma$)x+3x$^{2}$+x$^{3}$ = 0 with x = $\left(
{\frac{b}{a}}\right)^{2}$ = ${\delta}^{2}$ and $\gamma$ =
8(1+$\delta_1$J$^{2}$+$\delta_2$J$^{4}$)$^{3}$.
The quadrupole deformation parameter $\beta$ is equal to $\beta$ =
${\frac{1}{\sqrt{5\pi}}}({\frac{4}{3}}\delta+{\frac{2}{3}}\delta^{2}+{\frac{2}{3}}\delta^{3}+{\frac{11}{18}}\delta^{4})$.\\

The effects of the Yrast line lowering (increase of the level density) due to
the nuclear deformation and the variation of the deformation parameter $\beta$
can be quantitatively discussed using the values in Table \ref{table3} for
several reactions. The values of the minor to major axis ratio b/a and of the
deformation parameter $\beta$ have been extracted (with 10$\%$ error bars) from
the fitted deformability parameters by assuming a symmetric prolate shape with
sharp surfaces~\cite{Huizenga89}. The assumption of oblate shapes yields
similar results within 5$\%$. It is interesting to note that the deformation
found necessary to reproduce the $^{28}$Si + $^{12}$C reaction results is
smaller than the deformation introduced by the deformability parameter used by
Kildir {\it et al.}~\cite{Kildir95}, who also change the transmission
coefficients. 

The solid lines shown in Figures 6 and 7  are the results of {\sc CACARIZO}
calculations for the exclusive measurements. It can be observed in Fig.~6 that
for E$_{lab}$ = 112 MeV the experimental angular correlations are well
reproduced by the evaporation calculations for the data at the opposite side
from the ER detector, and this is true for correlations with both S and P ER's.
However the calculations fail to predict the experimental data at the same side
as the ER detector. The question of these large yields measured at negative
angles remains open~\cite{Bhattacharya02}. Similarly the {\sc CACARIZO}
calculations reproduce in Fig.~7 the in-plane angular correlations of $\alpha$
particles (circles) and protons in coincidence with all ER's, at E$_{lab}$ =
180 MeV, for the data on the opposite side from the ER detector. They are also
able to describe the in-plane angular correlations of protons in coincidence
with individual Z = 14 and Z = 15 on both sides of the beam. However, the
excess of yields observed at backward angles ($\Theta_{lab}$ = +50$^\circ$ to
+90$^\circ$) for $\alpha$ particles in coincidence with S may indicate the
occurrence of a non-evaporative process, possibly of a binary nature. The excess
of yields is even stronger for $\alpha$ particles in coincidence with Si. Here
the hypothesis of a non-evaporative component arising from the  $^{12}$C
breakup may be advanced. 

The solid lines shown in Fig.~8 for the out-of-plane angular correlations are
the results of {\sc CACARIZO} statistical-model predictions. Once again it can
be observed that the statistical-model calculations are able to reproduce the
proton coincidences well, but they fail to describe the $\alpha$-Cl
coincidences and the large yields found in coincidence with P and Si ER's in
the most forward direction. The poor reproduction of the $\alpha$-particle
experimental anisotropy factor is not well understood. The angular momentum
dependence has been tested by performing calculations with two different
angular-momentum windows : 10-20$\hbar$ and 20-30$\hbar$. Whereas for protons
the anisotropy is almost constant with the L-window, for the $\alpha$ particles
the anisotropy is strongly depending of the chosen L-window. Nevertheless the
flat behavior shown around 0$^\circ$ is present for the two particle species.
The observed discrepancy suggests that the  assumed angular distribution of the
LP, which is handled semi-classically , may not be adequate to describe the
out-of-plane data. 

Overall, we conclude that the evaporated $\alpha$ particles from $^{40}$Ca CN
emission reflect significant deformation effects. The deduced deformation is
comparable to that found previously in the analysis of
$^{28}$Si($^{12}$C,2$\alpha$)$^{32}$S$_{g.s.}$ angular correlation
data~\cite{Alamanos83}. The extent to which these effects can be reasonably
well quantified is dependent on the experimental coverage and, in particular,
on the power of the coincidence trigger. It is of particular interest to note
that the value of $\beta$ $\approx$ 0.5 found for the quadrupole deformation
parameter of $^{28}$Si + $^{12}$C (see Table III) might be connected with the
recent observation of SD bands in the doubly-magic $^{40}$Ca nucleus by
standard $\gamma$-ray spectroscopy methods~\cite{Ideguchi01}. Correlating large
prolate deformations in the hot CN with the presence of SD bands in $^{40}$Ca
is obviously not straightforward, since the deformation deduced from the LCP
data averages over CN configurations, while the SD bands are based on one of
these configurations. We made the same discussion with the possible comparison
between LCP results for the $^{28}$Si + $^{28}$Si
reaction~\cite{Bhattacharya02} and $\gamma$-ray data displaying very deformed
bands in the doubly magic $^{56}$Ni nucleus~\cite{Rudolph99}. It is interesting
to add that the macroscopic deformation energy of $^{40}$Ca recently calculated
within a Generalized Liquid Drop Model~\cite{Royer02} with shell effects (using
the Strutinsky method) generates a second highly deformed minimum where SD and
highly deformed states may survive. 

\subsection{Non-statistical $^{8}$Be cluster emission}

It has been shown from the analysis of Figs.~3 and 4 with {\sc CACARIZO} that
additional non-statistical components appear to be significant at both
E$_{lab}$ = 112 MeV and 180 MeV bombarding energies. However no evidence was
found for additional processes at the lower bombarding energies of E$_{lab}$ =
70 
MeV~\cite{Alamanos83} and 87 MeV~\cite{Ost80}. To better understand the origin
of these components, $\alpha$ particle energies are plotted in Fig.~9 against
the energies of the S residues detected at $\Theta_{S}$ = -10$^\circ$ for the
$^{28}$Si(180 MeV) + $^{12}$C reaction for a number of $\alpha$-particle
emission angles. With increasing $\alpha$-particle angles an increase of the
energy of S residues and a decrease of the $\alpha$ energy is observed which is
consistent with kinematics. At $\Theta_{\alpha}$ = +40$^\circ$, +45$^\circ$,
and +50$^\circ$ the bulk of events in Fig.~9 are of a statistical origin, and
consistent with {\sc CACARIZO} calculations, as demonstrated in Fig.~10-(b)
(for $\Theta_{\alpha}$ = +40$^\circ$). Another statistical-model code PACE 
2~\cite{Gavron80} gives similar predictions. The calculations suggest that
these $\alpha$ particles result from a cascade of a single $\alpha$, two
protons, and x neutrons rather than a 2-$\alpha$,xn evaporation process. For
larger angles, the two branches, corresponding to the contours labeled 1 and
2, although lying outside the ``statistical evaporation region'', still
correspond to an evaporation process as shown by the {\sc CACARIZO}
calculations displayed in Fig.~10-(b) for $\Theta_{\alpha}$ = +40$^\circ$ and
in Fig.~10-(d) for $\Theta_{\alpha}$ = +70$^\circ$. These two branches 1 and 2
correspond to a 2-$\alpha$ fusion-evaporation channel with both the $\alpha$
particles emitted respectively at backward and forward angles in the center of
mass. However, at more backward angles other additional contributions,
corresponding to the strong peak in the contour labeled 2 and in the contours
labeled 3 and 4, appear with increasing significance, as shown, for instance in
Fig.~10-(c) for $\Theta_{\alpha}$ = +70$^\circ$. The corresponding ``folding
angles'' are compatible with the two-body kinematics required for the $^{32}$S
+ $^{8}$Be binary exit-channel. In contrast, the energy correlations for the
$\alpha$ particles in coincidence with Cl and P residues (not shown) do not
exhibit similar two-body branches, the ``statistical evaporation region'' is
consistent with the {\sc CACARIZO} predictions, for all the measured angles. 

Although in principle the identification of the $^{8}$Be cluster requires the
coincident detection and mass identification of both decaying $\alpha$
particles~\cite{Mcdonald92}, a kinematic reconstruction assuming a two-body
$^{32}$S + $^{8}$Be$^{*}$ process is instructive.  Assuming the three-body
kinematics of a $\alpha$+$\alpha$+$^{32}$S final state, it is possible to
reconstruct the momentum of the ``missing'' $\alpha$-particle and, hence, to
deduce the excitation energy of the intermediate $^{8}$Be fragment. In Figs.~11
the deduced excitation energy spectra in this channel  are presented for the
contributions labeled 2, 3, and 4 in Fig.~10-(c) at the indicated
$\Theta_{\alpha}$ angles. From $\Theta_{\alpha}$ = 70$^\circ$ to
$\Theta_{\alpha}$ = 85$^\circ$ the strongest peak appears with a very narrow
width. This large component, which corresponds to the contribution of the
contour 4 visible in Fig.~10-(c), is centered at the energy of the ground state
of $^{8}$Be (the relative energy of the two $\alpha$ particles of the $^{8}$Be
breaking up in flight is 92 keV) and displayed as the squared part of
Fig.~11-(a). From $\Theta_{\alpha}$ = 55$^\circ$ to $\Theta_{\alpha}$ = 95
$^\circ$ the main bulk of the yields from contours 2 and 3 is centered at
around E$^{*}$ = 3.1 MeV with an experimental width of approximately 1.5 MeV,
which values correspond well to the known energy (E$^{*}$ = 3.04 MeV) and width
($\Gamma$ = 1500 keV) of the first 2$^+$ excited level of
$^{8}$Be~\cite{Ajzenberg88}. The short-lived $^{8}$Be 4$^+$ excited level at
E$^{*}$ = 11.4 MeV~\cite{Arena95} is not clearly observed due to its very broad
width ($\Gamma$ = 3.7 MeV) and the significant $\alpha$-statistical background
arising from the contribution of the contour 2. For the same reasons it is
hazardous to assign the bumps around 15 MeV to the known 2$^+$
doublet~\cite{Ajzenberg88} at E$^{*}$ = 16.6 and 16.9 MeV. 

Figs.~11-(b) and 11-(c) display the reconstructed excitation energy spectra of
the S binary fragments measured at $\Theta_{S}$ = -10$^\circ$ in coincidence
with $\alpha$ particles detected at the indicated $\Theta_{\alpha}$ angles by
gating either on the ground state (g.s.) contour 4 (upper panel) or the 2$^+$
state contours 2 and 3 (lower panel). We have performed fusion-fission
calculations (not shown), using the Extended Hauser-Feshbach Method
\cite{Matsuse97}. They fail to reproduce both the excitation energies of the
$^{32}$S fragments, and the yields from the contributions 2, 3 and 4
\cite{Rousseau00}. These contributions might result from a faster binary
process governed by the $\alpha$-transfer reaction mechanism $^{28}$Si +
$^{12}$C $\rightarrow$ $^{32}$S$^{*}$ + $^{8}$Be, as proposed by Morgenstern
{\it et al.}~\cite{Morgenstern86}. This conclusion is in agreement with
previous inclusive results published in Ref.~\cite{Arena94}. In the
cluster-transfer picture~\cite{Morgenstern86} the reaction is characterized by
a ``Q-value'' window centered at the so-called ``Q-optimum'', which value can
be estimated semi-classically by Q$_{opt}$ =
(Z$_{3}$Z$_{4}$/Z$_{1}$Z$_{2}$-1)E$_i^{c.m.}$, where the indices 1,2 and 3,4
indicate the entrance (i) and exit channel, respectively. The corresponding
excitation energy E$^{*}$ = Q$_{gg}$ - Q$_{opt}$, where Q$_{gg}$ is the
ground-state Q-value of the reaction. In this case the expected excitation
energy in the $^{32}$S nuclei is equal to 12.9 MeV. Fig.~11 represents the
calculated excitation energy of $^{32}$S in coincidence with the g.s. (b) for
individual angles and by adding individuals angles (lower spectrum labelled
Total), and with the first 2$^{+}$ (E$_x$ = 3.04 MeV) excited state (c) of
${^8}$Be, respectively. The strong shifts of the energy distributions can be
explained by the bias effects induced by the kinematic coincidence acceptances.
The dashed lines correspond to E$^{*}$ = 12.9 MeV, the energy expected for
$\alpha$-transfer reaction mechanisms. In both cases the excitation energies
(total spectra for the coincidence with the g.s.) of $^{32}$S are consistent
with these values. In the same way we can also have a $^{8}$Be-transfer
reaction mechanism \cite{Arena94} $^{28}$Si + $^{12}$C $\rightarrow$
$^{36}$Ar$^{*}$ + $\alpha$. In this case the $^{36}$Ar$^{*}$ ejectile has
sufficient excitation energy to emit either one proton or one $\alpha$
particle. This type of ``transferlike'' reaction can explain the disagreement
observed in Fig.~6 between data and {\sc CACARIZO} calculation for the in-plane
angular correlation between $\alpha$ particles and Cl residues. 

\bigskip

\section{CONCLUSIONS}

The possible occurrence of highly deformed configurations in the $^{40}$Ca
dinuclear system has been investigated by using the {\sc ICARE}
charged-particle multidetector array at the {\sc VIVITRON} Tandem facility of
the IReS Strasbourg. The properties of the emitted LCP's in the $^{28}$Si +
$^{12}$C reaction have been analyzed at two bombarding energies E$_{lab}$ =
112 MeV and 180 MeV, and compared with a statistical model that was adopted to
calculate evaporation spectra and angular distributions for deformed nuclei. A
Monte Carlo technique has been employed in the framework of the well documented
Hauser-Feshbach code {\sc CACARIZO}. The measured observables such as velocity
distributions, energy spectra, in-plane and out-of-plane angular correlations
are all reasonably well described by the Monte Carlo calculations which include
spin-dependent level densities. The magnitude of the adjustments in the Yrast
line suggests deformations at high spins in the $^{40}$Ca dinuclear system that
are far in excess of those predicted by the FRLDM. The deduced deformations are
comparable to recent $\gamma$-ray spectroscopy data for the $^{40}$Ca nucleus
at much lower spins~\cite{Ideguchi01}. The suggested $^{40}$Ca shapes are
consistent with predictions of the Generalized Liquid Drop Model~\cite{Royer02}
which predicts a second highly deformed minimum in this system resulting from
shell effects where SD and highly deformed states may survive. A component is
found in the $\alpha$-particle energy spectra measured in coincidence with S
residues that is attributed to the decay of unbound $^8$Be nuclei, although
this study does not clearly establish the mechanism resulting in these yields.
In general, to fully explore the influence of nuclear deformation on the
reaction mechanisms and underlying nuclear structure in the mass A$\approx$40
region will require sophisticated particle-$\gamma$ experiments (see Refs.
\cite{Nouicer99,Beck01,Thummerer00,Thummerer01} for instance) using {\sc
EUROBALL IV} and/or {\sc GAMMASPHERE}. These studies are necessary to better
understand how the large nuclear deformations that are apparent in the fusion
studies are related to the superdeformed bands discovered and/or predicted in
this mass region~\cite{Svensson00,Svensson01,Ideguchi01}. 

\bigskip

\centerline{\bf ACKNOWLEDGMENTS }

\vskip 0.8cm

\noindent
This paper is based upon the Ph.D.~thesis of M. Rousseau, Universit\'e Louis
Pasteur, Strasbourg, 2000. The authors would like to thank the staff of the
{\sc VIVITRON} for providing us with good stable beams, and J. Devin and C.
Fuchs for their excellent support in carrying out these experiments. Particular
appreciation to M.A. Saettel for preparing the targets, and to J.P.~Stockert
and A.~Pape for assistance during their RBS measurements. We also whish to
thank N. Rowley and W. Catford for useful discussions and for a careful reading
of the manuscript. One of us (M.R.) would like to acknowledge the Conseil
R\'egional d'Alsace for the financial support of his Ph.D.~thesis work. Parts
of this work has also been done in collaboration with C.E. during his summer
stay at the IReS with a JANUS Grant of IN2P3. This work was sponsored by the
French CNRS/IN2P3. Additional support was supplied by the U.S. DOE and the
CNRS/NSF and CNRS/CNPq International Collaboration Programs.

\renewcommand{\baselinestretch}{1.5}
\begin{table}[hptb]
\begin{tabular}{|c|c|c|c|c|c|}
\hline
\multicolumn{6}{|c|}{\sc ICARE setup} \\
\hline
\multicolumn{3}{|c|}{ $^{28}$Si(112 MeV) + $^{12}$C}  &  \multicolumn{3}{c|}{$^{28}$Si(180 MeV) + $^{12}$C }   \\ 
\hline
Type of detector & $\theta$($^\circ$) & $\phi$($^\circ$) & Type of detector & $\theta$($^\circ$) & $\phi$($^\circ$)\\
\hline
IC  & $\pm$15 & 0   & IC  & $\pm$10 & 0  \\
    & -20     & +20 &     & +10     & 90 \\
    & $\pm$25 & 0   &     & $\pm$15 & 0  \\
    & -30     & +20 &     & $\pm$20 & 0  \\
    & -35     & 0   &     & +20     & 90 \\
    & -40     & +20 &     & $\pm$25 & 0  \\
    & +130    & 0   &     &  -      &  -  \\
    & -150    & 0   &     &  -      &  -  \\
\hline
TL3 & +15     & +20 & TL3 & $\pm$30 & 0  \\
    & +25     & +20 &     & +35     & 0  \\
    & +35     & +20 &     & -       & -  \\
    & +45     & +20 &     & -       & -  \\
\hline
TL2 & \multicolumn{2}{c|}{ 16 telescopes each 5$^\circ$ from} & TL2 &
\multicolumn{2}{c|}{ 24 telescopes each 5$^\circ$ from}  \\ 
  & \multicolumn{2}{c|}{$\theta$=40$^\circ$ to 115$^\circ$ and
$\phi$=0$^\circ$} &     & \multicolumn{2}{c|}{$\theta$=40$^\circ$ to 95$^\circ$
and $\phi$=0$^\circ$} \\ 
\end{tabular}
\renewcommand{\baselinestretch}{1.}
\caption{\label{table1}\textit{\sl Experimental setup of {\sc ICARE} chosen for
the $^{28}$Si + $^{12}$C reaction at E$_{lab}$ = 112 MeV and 180 MeV.}} 
\end{table}
\renewcommand{\baselinestretch}{1.}

{\footnotesize
\begin{table}[htpb]
\begin{tabular}{|l|}
\hline
\hline
\centerline{\bf Angular-momentum distribution in CN}\\
\\
Critical angular momenta $L_{cr}$ = 21 (E$_{lab}$ = 112 MeV) and 27$\hbar$
(E$_{lab}$ = 180 MeV).\\
Diffuseness parameter $\Delta L = 1.0\hbar$.\\
\hline
\centerline{\bf OM potentials of the emitted LCP and neutrons}\\
\\
(1) Transmission coefficients as defined in the text\\
(2) Multiply factor of the OM radius: RFACT = 1\\
\hline
\centerline{\parbox{14cm}{\bf Level-density parameters at low excitation:
(E$^* \leq$ 10 MeV)}}\\
\\
(1) \parbox{15cm}{Fermi-gas level-density formula with empirical level-density
parameters from Dilg {\it et al.}~\cite{Dilg73}}\\
(2) Effective moment of inertia $\Im$ = IFACT $\Im_{rigid}$ with IFACT = 1.\\
\hline
\centerline{\parbox{14cm}{\bf Level-density parameters at high excitation:
(E$^* \geq$ 20 MeV)}}\\
\\
(1) \parbox{15cm}{ Fermi-gas level-density formula with parameters from RLDM
(Myers and Swiatecki~\cite{Myers66})}\\
(2) Level-density parameter: $a$ = A/8 MeV$^{-1}$\\
\hline
\centerline{\bf Yrast line}\\
\\
- Parameter set {\bf A}: FRLDM (Sierk~\cite{Sierk86})
\\
- Parameter set {\bf B}: $\Im = \Im_{sphere} (1 + \delta_1 J^2 +
\delta_2 J^4)$ with $\delta_1$ = 2.5 10$^{-4}$ et $\delta_2$ = 5.0 10$^{-7}$\\
\hline
\centerline{{\bf $\gamma$-ray width} (in Weisskopf units)}\\
\\
(1) E1: B(E1) = 0.001\\
(2) M1: B(M1) = 0.01\\
(3) E2: B(E2) = 5.0\\
\hline

\end{tabular}

\caption{\label{table2}\textit{\sl Parameter sets used in the CACARIZO
calculations for the $^{28}$Si + $^{12}$C reaction at E$_{lab}$ = 112 and 180
MeV.}}

\end{table}}

\renewcommand{\baselinestretch}{1.2}
\begin{table}[h!]
\hspace{-2cm}
\begin{tabular}{|c|c|c|c|c|c|c|c|}
\hline
Reaction & C.N. & $L_{cr}$ ($\hbar$) & $\delta_1$ & $\delta_2$ &
b/a & $\beta$ & Reference\\ 
\hline
$^{28}$Si + $^{12}$C & $^{40}$Ca  & 21 & 2.5$\cdot$10$^{-4}$ &
5.0$\cdot$10$^{-7}$ & 1.3 & 0.47 & This work\\ 
\hline
$^{28}$Si + $^{12}$C & $^{40}$Ca  & 26 & 6.5$\cdot$10$^{-4}$ &
3.3$\cdot$10$^{-7}$ & 2.0 & 0.53 & \cite{Kildir95} \\
\hline
$^{28}$Si + $^{12}$C & $^{40}$Ca  & 27 & 2.5$\cdot$10$^{-4}$ &
5.0$\cdot$10$^{-7}$ & 1.8 & 0.51 & This work\\ 
\hline
$^{28}$Si + $^{27}$Al & $^{55}$Co & 42 & 1.8$\cdot$10$^{-4}$ &
1.8$\cdot$10$^{-7}$ & 1.3 & 0.46 & \cite{Agnihotri93}\\
\hline
$^{28}$Si + $^{28}$Si & $^{56}$Ni & 34 & 1.2$\cdot$10$^{-4}$ &
1.1$\cdot$10$^{-7}$ & 1.6 & 0.49 & \cite{Bhattacharya02}\\ 
\hline
$^{28}$Si + $^{28}$Si & $^{56}$Ni & 37 & 1.2$\cdot$10$^{-4}$ &
1.1$\cdot$10$^{-7}$ & 1.7 & 0.50 & \cite{Rousseau01a}\\
\hline
$^{30}$Si + $^{30}$Si & $^{60}$Ni & 34 & 1.2$\cdot$10$^{-4}$ &
1.1$\cdot$10$^{-7}$ & 1.7 & 0.50 & \cite{Bhattacharya02}\\ 
\hline
$^{35}$Cl + $^{24}$Mg & $^{59}$Cu & 37 & 1.1$\cdot$10$^{-4}$ &
1.3$\cdot$10$^{-7}$ & 1.7 & 0.51 & \cite{Mahboub02}\\  
\hline
$^{32}$S + $^{27}$Al & $^{59}$Cu & 42 & 1.3$\cdot$10$^{-4}$
& 1.2$\cdot$10$^{-7}$ & 2.0 & 0.53 & \cite{Huizenga89}\\  
\hline
$^{16}$O + $^{54}$Fe & $^{70}$Se & 34 & 2.5$\cdot$10$^{-5}$ &
3.0$\cdot$10$^{-8}$ & 1.3 & 0.46 & \cite{Govil00a}\\
\end{tabular}
\renewcommand{\baselinestretch}{1.}
\caption{\label{table3}\textit{ \sl Typical quantities of the evaporation
calculations performed using the statistical-model code {\sc CACARIZO} as
discussed in the text. The deformability parameters are taken either from the
parameter {\bf set B} (see Table II) for $^{28}$Si + $^{12}$C or from similar
fitting procedures for the other systems studied in the literature. }} 
\end{table}

\begin{figure}
Figure 1: Experimental C (solid squares), N (solid triangles), and O (solid
circles) cross sections measured in the $^{28}$Si + $^{12}$C reaction
\cite{Shapira82,Shapira84} as compared to the calculations (dotted curves)
performed with the equilibrium model of orbiting \cite{Shivakumar87}. The
solid curves are the predictions of the transition-state model
\cite{Sanders99}. The open squares, triangles and circles are the present data
of the C, N, and O fully-damped yields with error bars smaller than the size
of the symbols. The full diamonds correspond to ER cross sections quoted in
Refs.~\cite{Gary82,Lesko82,Nagashima82,Harmon86,Harmon88,Vineyard93,Arena94}. 
\end{figure}

\begin{figure}
Figure 2:  Inclusive energy spectra of $\alpha$ particles measured in the
$^{28}$Si + $^{12}$C reaction at E$_{lab}$ = 112 MeV (a) and (b), and 180 MeV
(c) and (d) at $\Theta^{LCP}_{lab}$ = 40$^\circ$ and 45$^\circ$. The
experimental data are shown by the solid points with error bars visible when
greater than the size of the points. The solid and dashed lines are
statistical-model calculations discussed in the text. 
\end{figure}

\begin{figure}
Figure 3: Exclusive energy spectra of $\alpha$ particles emitted at the angles
+40$^{\circ} < \Theta^{LCP}_{lab} < ~$+65$^{\circ}$, in coincidence with
individual P and S ER's detected at -15$^\circ$ in the $^{28}$Si + $^{12}$C
reaction at E$_{lab}$ = 112 MeV. The experimental data are given by the solid
points with error bars visible when greater than the size of the points. The
solid lines are statistical-model calculations discussed in the
text. 
\end{figure}
        
\begin{figure}
Figure 4 Exclusive energy spectra of $\alpha$ particles emitted at the angles
+40$^{\circ} < \Theta^{LCP}_{lab} < ~$+95$^{\circ}$, in coincidence with
individual P and S ER's detected at -10$^\circ$ in the $^{28}$Si + $^{12}$C
reaction at E$_{lab}$ = 180 MeV. The experimental data are given by the solid
points with error bars visible when greater than the size of the points. The
solid lines are statistical-model calculations discussed in the
text. 
\end{figure}
        
\begin{figure}
Figure 5: Exclusive energy spectra of protons emitted at the angles
+40$^{\circ} < \Theta^{LCP}_{lab} < $+70$^{\circ}$, in coincidence with
individual P and S ER's detected at -10$^\circ$, at the indicated laboratory
angles, in the $^{28}$Si(180 MeV) + $^{12}$C reaction. The solid lines are
statistical-model calculations discussed in the text. 
\end{figure}

\begin{figure}
Figure 6: In-plane angular
correlations of $\alpha$ particles (circles) and protons (triangles) measured
in coincidence with the ER's Z = 16 (a) and 15 (b) in the $^{28}$Si + $^{12}$C
reaction at E$_{lab}$ = 112 MeV in the angular range -115$^{\circ}$ $\leq$
$\Theta^{LCP}_{lab}$ $\leq$ +115$^{\circ}$). The proton correlations have been
multiplied by a factor 10$^{-3}$ for the sake of clarity. The arrow indicates
the position of the IC detector at $\Theta_{lab}$= -15$^\circ$. On the
abscissa, the positive angle refer to the opposite side of the beam from the
direction of the ER detected in IC. The solid lines correspond to
statistical-model calculations discussed in the text. 
\end{figure}

\begin{figure}
Figure 7: In-plane angular correlations of $\alpha$ particles (circles) and
protons (triangles) measured in coincidence with the ER's with Z = 17 (a), 16
(b), 15 (c), and 14 (d) in the $^{28}$Si + $^{12}$C reaction at E$_{lab}$ = 180
MeV. in the angular range -115$^{\circ}$ $\leq$ $\Theta^{LCP}_{lab}$ $\leq$
+115$^{\circ}$). The proton correlations have been multiplied by a factor
10$^{-2}$ for the sake of clarity. The arrow indicates the position of the IC
detector at $\Theta_{lab}$= -10$^\circ$. On the abscissa, the positive angle
refer to the opposite side of the beam from the direction of the ER detected in
IC. The solid lines correspond to statistical-model calculations discussed in
the text. 
\end{figure} 

\begin{figure}
Figure 8: Out-of-plane angular correlations of coincident $\alpha$ particles
(circles) and protons (triangles) measured in the $^{28}$Si + $^{12}$C reaction
at E$_{lab}$ = 180 MeV. The proton correlations have been multiplied by a
factor 10$^{-2}$ for the sake of clarity. The ER's with Z = 17 (a), 16 (b), 15
(c), and 14 (d) are detected at $\Theta_{lab}$ = -10$^\circ$. The solid lines
correspond to statistical-model calculations discussed in the text. 
\end{figure}

\begin{figure}
Figure 9: Energy-correlation plots between coincident $\alpha$ particles and S
ER's produced in the $^{28}$Si + $^{12}$C reaction at E$_{lab}$ = 180 MeV. The
heavy fragment is detected at $\Theta_{S}$ = -10$^\circ$ and the
$\alpha$-particle angle settings are given in the figure. The dashed lines
correspond to different contours with their associated labellings discussed in
the text. 
\end{figure}

\begin{figure}
Figure 10: Experimental (a,c) and calculated (b,d) energy-correlation plots
between coincident $\alpha$ particles and S ER's produced in the $^{28}$Si +
$^{12}$C reaction at E$_{lab}$ = 180 MeV. The S is identified at $\Theta_{S}$ =
-10$^\circ$ and the $\alpha$ particles are detected at $\Theta_{\alpha}$ =
+40$^\circ$, and +70$^\circ$, respectively. {\sc CACARIZO} calculations are
discussed in the text. 
\end{figure}

\begin{figure}
Figure 11: Excitation-energy spectra calculated for the $^{28}$Si + $^{12}$C
reaction at E$_{lab}$ = 180 MeV for $^{8}$Be (a) and $^{32}$S in coincidence
with the g.s.(b) and first excited level (c) of $^{8}$Be. The solid line
corresponds to the energy of the first excited state of $^{8}$Be (3.08 MeV).
The dashed lines correspond to an excitation energy in $^{32}$S expected for an
$\alpha$-transfer process. The lower spectrum in (b) corresponds to the
sum of the 4 individuals spectra. 
\end{figure}

\end{document}